\documentclass[12pt,reqno]{amsart}
\usepackage{amsfonts, amsmath, amssymb, amsthm, amsbsy}

\include{diagrams}
\diagramstyle[small,scriptlabels,
midhshaft,nohug]
\newarrow{TeXto}----{->}
\newcommand{\rto}{\rTeXto}
\newcommand{\lto}{\lTeXto}
\newcommand{\uto}{\uTeXto}
\newcommand{\dto}{\dTeXto}

\newcommand{\rdto}{\rdTeXto}
\newcommand{\luto}{\luTeXto}

\newtheorem{lemma}{Lemma}

\newtheorem{state}{Proposition}
\theoremstyle{remark}
\newtheorem{rem}{Remark}

\newcommand{\Id}{{\mathrm d}}

\DeclareFontFamily{OML}{cyr}{}
\DeclareFontShape{OML}{cyr}{m}{n}{
   <5> <6> <7> <8> <9> gen * wncyr
   <10> <10.95> <12> <14.4> <17.28> <20.74> <24.88> wncyr10
  }{}
\DeclareSymbolFont{rusletters}{OML}{cyr}{m}{n}
\DeclareSymbolFontAlphabet{\rusmath}{rusletters}
\DeclareMathSymbol\re{\rusmath}{rusletters}{"03}
\newcommand{\cEv}{\re}

\newcommand{\BBC}{{\mathbb{C}}}
\newcommand{\BBR}{{\mathbb{R}}}
\newcommand{\sym}{\mathop{\rm sym}\nolimits}
\newcommand{\const}{\mathop{\rm const}\nolimits}

\newcommand{\gm}{\mathfrak{m}}
\newcommand{\bgm}{{\bar{\mathfrak{m}}}}

\newcommand{\bu}{{\boldsymbol{u}}}
\newcommand{\bU}{{\boldsymbol{U}}}
\newcommand{\ba}{{\boldsymbol{a}}}
\newcommand{\balpha}{{\boldsymbol{\alpha}}}

\newcommand{\bE}{\mathbf{E}}

\newcommand{\bw}{{\boldsymbol{W}}}
\newcommand{\gA}{\mathfrak{A}}
\newcommand{\gB}{\mathfrak{B}}

\newcommand{\cE}{\mathcal{E}}
\newcommand{\cEEL}{{\mathcal{E}}_{\text{\textup{EL}}}}
\newcommand{\cH}{\mathcal{H}}
\newcommand{\cL}{\mathcal{L}}
\newcommand{\dd}{\partial}
\newcommand{\bun}{\mathbf{1}}
\newcommand{\vph}{\varphi}
\newcommand{\vth}{\vartheta}

\newcommand{\pKdV}{{\text{\textup{pKdV}}}}

\newcommand{\KdV}{{\text{\textup{KdV}}}}
\newcommand{\mKdV}{{\text{\textup{mKdV}}}}
\newcommand{\Toda}{{\text{\textup{Toda}}}}
\newcommand{\Bous}{{\text{\textup{Bous}}}}

\newcommand{\pp}{\phantom{+}}

\newcommand{\by}[1]{\textit{{#1}}}
\newcommand{\jour}[1]{\textit{{#1}}}
\newcommand{\vol}[1]{\textbf{{#1}}}
\newcommand{\book}[1]{\textrm{{#1}}}

\begin{document}
\title{On Hamiltonian flows on Euler-type equations}

\date{December 22, 2004}

\author{A.\,V.\,Kiselev}

\thanks{Supported by INTAS grant
YS\,2001/2-33 and University of Lecce grant No.\,650\,CP/D}

\thanks{Submitted to \textit{Theoretical \& Mathematical Physics},
Proc.\ conf.\ ``Nonlinear Physics: Theory and Experiment~III''
(Gallipoli, 2004)}

\address{
\textup{\textrm{Permanent address:}}
153003 Russia, Ivanovo, Rabfakovskaya str.\,34,
Ivanovo State Power University, Dept.\ of Higher Mathematics.}

\curraddr{
Department of Mathematics, Brock University, 500 Glenridge Ave.,
St.~Catharines, Ontario, Canada L2S~3A1.}

\email{arthemy.kiselev@brocku.ca}



\rightline{ISPUmath-3/2004 ver.\,2}


\begin{abstract}
Properties of Hamiltonian symmetry flows on
hyperbolic Euler\/-\/type Liouvillean equations~$\cEEL'$ are analyzed.
Description of their Noether symmetries assigned to the integrals
for these equations is obtained.
The integrals provide Miura transformations from~$\cEEL'$ to the
multi\/-\/component wave equations~$\cE$.
By using these substitutions, we generate an infinite\/-\/Hamiltonian
commutative subalgebra~$\gA$ of local Noether symmetry flows on~$\cE$
proliferated by weakly nonlocal recursion operators. We demonstrate that
the correlation between the Magri schemes for~$\gA$ and for the induced
``modified'' Hamiltonian flows~$\gB\subset\sym\cEEL'$ is such that
these properties are transferred to~$\gB$ and
the recursions for~$\cEEL'$ are factorized.
Two examples associated with the 2D~Toda lattice are considered.
\end{abstract}

\maketitle

\subsection*{Introduction.}
In this paper, we consider the problem of constructing pairs of
commutative hierarchies of Hamiltonian evolution equations related by
Miura-type transformations and identified with Lie subalgebras of the
Noether symmetry algebras for Euler\/--\/Lagrange\/-\/type systems.
By using two standard schemes (\cite{SokolovUMN, Magri, Miura}),
which are
the Miura substitutions defined by the integrals of Liouvillean
hyperbolic equations and construction of the second Hamiltonian
structure by a Miura transformation, we restrict the exposition to the
class of the Euler\/--\/Lagrange Liouvillean hyperbolic systems~$\cEEL'$.
We obtain an explicit description of their Noether symmetries assigned
to the integrals for~$\cEEL'$; these flows are Hamiltonian w.r.t.\ the
(first) structures derived from the Lagrangians.
Then we analyze the properties of the Magri schemes for two
hierarchies:~$\gA$,
which is composed by symmetries of the wave equation,
and~$\gB\subset\sym\cEEL'$, correlated by the Miura maps.
Two examples are discussed.
First, we relate the Korteweg\/--\/de Vries equation
\begin{equation}\label{BothKdV}
s_{t_1}=-\beta\,s_{xxx}+\tfrac{3}{2}s_x^2,\quad
w_{t_1}=-\beta\,w_{xxx}+3ww_x,\quad w=s_x,\ \beta=\const,
\end{equation}
and multi\/-\/component modified KdV
equations (see~\cite{DynamSys}) with the wave equation and
the two\/-\/dimensional Toda lattice
(2DTL, in particular, associated with a semisimple Lie
algebra, see~\cite{LeznovSmirnovShabat}), respectively,
and factorize the Lenard recursion operator for
Eq.~\eqref{BothKdV} to a product of two vector\/-\/valued operators.
Second, we demonstrate that
the Boussinesq equation~(\cite{FordyGibbons})
\begin{equation}\label{BothBous}
\left\{\begin{aligned}
u_{t_1}&=\tfrac{1}{3}v_{xxx}+\tfrac{4}{3}v_x^2 \\ v_{t_1}&=u_x
\end{aligned}
\right.\qquad
\left\{\begin{aligned}
U_{t_1}&=V_x \\ V_{t_1}&=\tfrac{1}{3}U_{xxx}+\tfrac{8}{3}UU_x
\end{aligned}
\right.\qquad
\left\{\begin{aligned}
U&=v_x \\ V&=u_x
\end{aligned}
\right.
\end{equation}
and the modified Boussinesq equations (\cite{PavlovFPM}), as well as the
Hamiltonian structures for their local commutative hierarchies, are
obtained from the geometry of the ambient wave and 2DTL equations,
respectively.

The relationship between the Hamiltonian and Lagrangian approaches
towards integrable evolution equations was discussed
in~\cite{NutkuPavlov}, where Lagrangian representations were derived by
using the Legendre transform from the sequence of Hamiltonian
functionals for an evolution equation at hand. Therefore, that concept
was closed w.r.t.\ evolutionary systems
and the Miura\/-\/type transformations between them, whichever
representation it might be. Our approach is opposite
to~\cite{NutkuPavlov}. Indeed, we interpret
(bi-)Hamiltonian hierarchies of evolution equations
as flows defined by subalgebras of the Noether symmetry algebras
for ambient hyperbolic
Euler\/-\/type systems (see also~\cite{FordyGibbons}).
By using the canonical variables for the Euler\/-\/type systems
and treating the differential constraint between coordinates $u$ and
momenta $\gm$ as the rule that defines the Clebsch potentials, we establish
the relation between the potential and nonpotential components of the
hierarchies that
describe the evolution of coordinates and momenta, respectively.
The adjoint linearization $(\ell_\gm^u)^*$
of this constraint defines the first Hamiltonian
structures for these hierarchies such that their Magri schemes are
correlated; the second Hamiltonian structures
obtained from the Miura transformations
supply the recursion
operators both for the evolutionary hierarchies and the ambient
Euler\/-\/type systems.   

The paper is organized as follows. In Sec.~\ref{SecTheor}, we relate
the differential
constraint between coordinates and momenta for hyperbolic Euler
equations with the Hamiltonian operators for their
Noether symmetry algebras.
Then we consider substitutions
from Euler\/-\/type Liouvillean equations~$\cEEL'$
generated by their integrals;\ in this case,\ the Miura transformation to a
subalgebra $\gA$ of the Noether symmetries of the Euler
wave equation~$\cE$
generates the second Hamiltonian structure on~$\gA$.
Thus we obtain the pair~$\gA$, $\gB$ of sequences of Hamiltonian flows
on~$\cE$ and $\cEEL'$, respectively, which are correlated by the Miura map.
Then we discuss properties of the flows within~$\gA$ and $\gB$
and the corresponding recursions.
Finally,\ we demonstrate which of these properties for~$\gA$
such as the locality,\ commutativity,\ (bi-)Ha\-mil\-to\-ni\-a\-ni\-ty, etc.,
are transferred to~$\gB$.

In Sec.~\ref{SecKdV}, we recall necessary facts about geometry of the
2DTL $u_{xy}=\exp(Ku)$;
all notions and notation follow~\cite{ClassSymEng}.
We consider the hierarchy of KdV equation~\eqref{BothKdV}
and the sequence of multi\/-\/component analogues (\cite{DynamSys})
of the modified KdV equations associated with the~2DTL.
We obtain a factorization of the Lenard recursion operator for KdV
to a product of vector\/-\/valued operators; also, we prove that the
higher flows for the multi\/-\/component mKdV are local and pairwise
commute.
In Sec.~\ref{SecBous}, we demonstrate that
the hierarchy~$\gB$ of higher flows for modified
Boussinesq equation~\eqref{mBous} is a local commutative subalgebra
of Noether symmetries of 2DTL~\eqref{TodaFixedPoint} associated with
the algebra~$\mathfrak{sl}_3(\BBC)$ such that
integrals~\eqref{IntegralsmBous} for Eq.~\eqref{TodaFixedPoint}
induce the second Hamiltonian structure~\eqref{BousSecondHam} for
Eq.~\eqref{BothBous}; a factorization of the recursion operator for
2DTL is obtained.

\section{Hierarchies and Miura transformations}%
\label{SecTheor}
Consider a first\/-\/order Lagrangian
$\cL=[L(u, u_x, u_y; x, y)\,\Id x\wedge\Id y]$
with the density
$L=-\tfrac{1}{2}\sum\nolimits_{i,j}\bar\kappa_{ij}u^i_xu^j_y+H(u;
x, y)$, where $\bar\kappa$ is a real, nondegenerate, symmetric,
constant $(r\times r)$-matrix. Choose the 
variable $y$ for the ``time'' coordinate,
leave $x$ for the spatial coordinate,
and denote by $\gm_j=\dd L/\dd u^j_y$ the $j$th conjugate coordinate
(momentum) for the $j$th dependent variable $u^j$ for any $1\leq j\leq r$:
\begin{equation}\label{Constraint}
\gm=-\tfrac{1}{2}(\bar\kappa u_x)^*.
\end{equation}
We claim that the adjoint inverse linearization
$\smash{D_x^{-1}\circ\bar\kappa^{-1}}$ of
differential constraint~\eqref{Constraint} between the coordinates and
momenta is a Hamiltonian structure for the hyperbolic Euler equation
$\cEEL=\{\bE_u(\cL)=0\}$ itself and also for the algebra~$\sym\cL$ of
Noether symmetries for~$\cEEL$.
Consider the Legendre transform
$H\,\Id x\wedge\Id y=
\left\langle\gm,{\dd L}/{\dd u_y}\right\rangle - \cL$
and hence assign the Hamiltonian
$\cH(u$, $\gm)=[H\,\Id x]$ to the Lagrangian~$\cL$.
The hyperbolic Euler equation $\cEEL$ is equivalent to the system
$u_y=\bun\cdot\bE_\gm{(\cH)}$, $\gm_y=-\bun\cdot\bE_u{(\cH)}$
that involves the {canonical} Hamiltonian
structure~$\left(\begin{smallmatrix}\pp0 & \bun\\
 -\bun & 0\end{smallmatrix}\right)$.
Owing to the relations
$
\tfrac{1}{2} \bE_\gm=
{\bar\kappa^{-1}\cdot D_x^{-1}}\cdot \bE_u$, $
\bE_u=\tfrac{1}{2}
{D_x\circ\bar\kappa}\cdot \bE_\gm$,
the dynamical equations are separated. Indeed, they are evolutionary:
\label{eqLagHamForm}
$
u_y=A_1\circ\bE_u\bigl([H[u]\,\Id x]\bigr)$, $
\gm_y=-\tfrac{1}{2}\smash{\Hat A_1}\circ\bE_\gm\bigl([H[\gm]\,\Id x]\bigr)
$, 
that is, they are obtained by using
the pair of mutually inverse Hamiltonian operators
$A_1=\bar\kappa^{-1}\cdot D_x^{-1}$
and~$\smash{\hat A_1}=D_x\cdot\bar\kappa$.

Consider a hyperbolic Euler equation $\cEEL$ and suppose that it
admits a symmetry flow $u_t=\phi(u_x$, $u_{xx}$, $\ldots)\equiv\phi[u_x]$;
here~$t$ is a parameter along the integral trajectories. Then the
evolution $\gm_t$ of the momenta is described by the nonpotential
equation $\gm_t=-\tfrac{1}{2} \bar\kappa\cdot
D_x\bigl({}^t\phi[\gm]\bigr)$; see
Eq.~\eqref{BothKdV} for an example of symmetry flows on the wave
equation~$s_{xy}=0$.

\begin{lemma}[\textup{\cite{MeshkovTMPh}}]\label{MeshkovGenF}
Let $\cE=\{\bu_{xy}=\boldsymbol{f}(\bu;x,y)\}$ be a quasilinear
hyperbolic equation and $\cH=H\,\Id x+\cdots$ be its conservation
law\textup{:} $\Id_h(\cH)=\nabla(\bu_{xy}-\boldsymbol{f})$.
Then the \emph{generating section} $\psi\equiv\nabla^*(1)$ of the
conservation law~$\cH$ is~$\psi=-D_x(\bE_u(H\,\Id x))$.
\end{lemma}

By the Noether theorem~(\cite{Mystique}), the Noether symmetries
for self\/-\/adjoint Euler systems $\cEEL=\{\bE_u(\cL)=0\}$
coincide with the generating sections of their conservation laws;
by~\cite{TodaLawsActa}, the correlation between
the Noether symmetries~$\vph_\cL$ and the generating sections~$\psi$
for nonselfadjoint Euler equations
$\cEEL\simeq\{\bar\kappa^{-1}\cdot\bE_u(\cL)=0\}$ with a unit symbol
is~$\psi=\bar\kappa\vph_\cL$.

Let~$\phi$ be a Noether flow on an Euler\/-\/type hyperbolic system:
$u_t=\tfrac{1}{2}\, \bE_\gm(\cH)$,
$\gm_t=-\tfrac{1}{2}\, \bE_u(\cH)$, where $\cH=[H\,\Id x]$
and~$H$ is a conserved density. Hence we have
$u_t=\bar\kappa^{-1} \cdot D_x^{-1} \,\bE_u(\cH)$,
$\gm_t=-\tfrac{1}{2} D_x\cdot\bar\kappa\,\bE_\gm(\cH)$,
{i.e.}, both equations
$u_t=\phi$ and $\gm_t=-\smash{\tfrac{1}{2}}\bar\kappa\,D_x({}^t\phi)$
are Hamitonian simultaneously and their Hamiltonian operators
$A_1$ and $\smash{\Hat A_1}$ are mutually inverse
(e.g.,~\cite{Gardner}).
The induced evolution of momenta $\gm$ can be calculated in
two distinct ways: by variation of the Hamiltonian,
$\gm_t=-\bE_u(\cH)=-(\ell_\gm^u)^*(\bE_\gm(\cH))$, or by
using relation~\eqref{Constraint} explicitly:
$\gm_t=\ell_\gm^u(\bE_\gm(\cH))$.
Two evolutions are correlated for hyperbolic Euler equations since
the condition
$(\ell_\gm^u)^*=-\ell_\gm^u$
holds.
Further, let $R$ be a recursion for the hyperbolic Euler
equation $\cEEL=\{\bE_u(\cL)=0\}$ such that $R$ generates the sequence
$u_{t_i}=\phi_i[u_x]$ composed by its commuting
Hamiltonian symmetries originating from some $\phi_{-1}$.
The recursion $R$ is shared by all equations
$u_{t_i}=\phi_i$; the operator $R^*$ maps the velocities
$\gm_{t_i}$ of evolution of the momenta\,(see\,\cite{Lstar}).

From now on, we investigate a particular class of \emph{Liouvillean}
hyperbolic Euler equations
(\cite{SokolovUMN, LeznovSmirnovShabat, ShabatYamilov})
which admit nontrivial
\emph{integrals}, that is, functionals $w\in\ker\bar D_y$ belonging to
the kernel of the total derivative $D_y$ restricted onto
the equation~$\cEEL'$ at hand. Our reasonings hold up to
the involution~$x\leftrightarrow y$ if it is a symmetry of~$\cEEL'$,
otherwise the exposition admits the mirror copy with $x$ and $y$
replaced.

Suppose that the integrals $w[\bgm]$ depend on the
momenta $\bgm$ for a Liouvillean Euler-type equation.
Then the equation itself and its symmetry algebra are mapped
by the Miura transformation $\gm=w[\bgm]$ to the
multi\/-\/component wave equation~$\cE$ and its symmetry algebra,
respectively. The number of components within the wave equation equals
the number of the integrals~$w$ involved in the Miura map.
Also, two other mappings are well defined: the
covariant mapping $w_*\colon\sym\cEEL'\to\sym\cE$ of the symmetry flows
(see Proposition~4 in~\cite{SokolovUMN}) and the
contravariant mapping $w^*$ of the
Hamiltonians $H\,\Id x$ (and hence, of the Hamiltonian flows).
Indeed, by the transformation $\gm=w[\bgm]$, the
variational derivatives w.r.t.\ the momenta $\gm$ and
the modified momenta $\bgm$
are correlated by the rule $\bE_\bgm=\square\circ\bE_\gm$,
where $\square=(\ell_\gm^\bgm)^*$.
Hence from Lemma~\ref{MeshkovGenF} we deduce

\begin{state}\label{NoetherForm}
The adjoint linearization $\square=(\ell_w^\bgm)^*$
of the integrals~$w$ for a Liouvillean
Euler\/-\/type equation $\cEEL'=\{\bE_{\bar u}(\cL')=0\}$
w.r.t.\ its momenta $\bgm$ factors
its Noether symmetries\textup{:}
$\vph_{\cL'}=\square(\bE_\gm(\cH[\gm]))$, where~$\cH$ is arbitrary.
\end{state}

Examples are found in~\cite{TodaLawsActa, DynamSys}.
Recently, the correlation between the structure of generators of the
symmetry algebra for Liouvillean hyperbolic systems and their integrals
was analyzed in~\cite{Demskoi}.

Consider the substitution $\gm=w[\bgm]$ between an Euler\/-\/type
Liouvillean system~$\cEEL'$ and the multi\/-\/component wave
equation~$\cE$. The pair~$(w^*$, $w_*)$ generates the
second Hamiltonian structure $\smash{\hat A_2}$ for Noether symmetries of the
target equation~$\cE$, see~\cite{Miura}.
Namely, let~$\gA$ be a sequence of Noether flows on~$\cE$ such that
$H_i[\gm]$ are the (Hamiltonian) conserved densities for the first
structures $((\ell_\gm^u)^*)^{\pm1}$. 
Thence for any Hamiltonian flow $\phi_{i-1}=\bE_\gm(\cH_i)$ on~$\cE$
we obtain the flow $\bar u_{t_i}=\vph_i=\square(\phi_{i-1})$ on~$\cEEL'$.
Now use the condition $(\ell_\gm^u)^*=-\ell_\gm^u$, 
hence the evolution of the
modified momenta $\bgm$ is well defined.
The evolution $\smash{\cEv_{\square(\phi_{i-1})}(w)}$ of
the substitution $w$ along the flow
$\vph_i\in\gB$ is correlated with the flow $\phi_i$ that succeeds
$\phi_{i-1}$ in $\gA$ if
the second Hamiltonian structure~$\smash{\hat A_2}$ in~$\gA$ satisfies
the operator equation
\begin{equation}\label{eq4parts}
\ell_\gm^\bgm\circ(\ell_\bgm^{\bar u})^*\circ(\ell_\gm^\bgm)^*=
\smash{\hat A_2};
\end{equation}
and conversely, the operator~$\smash{\hat A_2}$ defined in~\eqref{eq4parts}
is always a Hamiltonian structure for~$\gA$.
In other words,
Eq.~\eqref{eq4parts} is the condition of reducibility of the structure
$\smash{\hat A_2}$ to the canonical form ``$\Id/\Id x$''.
Condition~\eqref{eq4parts} specifies
the set of admissible Miura transformations $w=w[\bgm]$ and modified
Euler equations $\cEEL'$ for a given~$\smash{\hat A_2}$.

\begin{rem}\label{RemInfinity}
If $\smash{\hat A_2}$ is compatible with~$\smash{\hat A_1}$,
then~$\gA$ is commutative and therefore the recursion
$R=\smash{(\hat A_2\circ\hat A_1^{-1})^*}$ for\,$\cE$ is
a recursion for\,$\gA$.
Thence, the bi\/-\/Ha\-mil\-to\-ni\-an
hierarchy~$\gA$ is infinite\/-\/Hamiltonian.\ %
Namely,\ let $\varepsilon=\{\bu_t=\smash{\hat A_1}(\psi[\bu])\}$
be a Hamiltonian equation and let $R$ be a recursion for~$\varepsilon$;
in our case, the operator~$R$ proliferates the commuting symmetry flows
on the wave equation~$\cE$.
Then $\varepsilon$ is also Hamiltonian w.r.t.\ the operators
$\smash{\hat A_i}=R^{i-1}\hat A_1$
provided that they are skew\/-\/adjoint;
it is indeed so for Eq.~\eqref{eq4parts}.
The proof is based on the following argument.
By~\cite{Lstar, Olver},
the relation $\ell_\varepsilon A + A\ell_\varepsilon^*=0$
holds for Hamiltonian operators~$A$ for~$\varepsilon$;
the converse is also true under regularity assumptions.
Since $R$ is a recursion for $\varepsilon$, we have
$\ell_\varepsilon R = R \ell_\varepsilon$. Indeed, from
the operator equality $\ell_\varepsilon R = \bar R \ell_\varepsilon$
that holds for some~$\bar R$, we obtain
$\bar R=R$ by commuting this equality with~$t$. Therefore, the identity
$\ell_\varepsilon (RA) + (RA) \ell_\varepsilon^*=0$ holds
and thus the operator $RA$ is also
Hamiltonian. Now proceed by induction. These reasonings are rigorous for
those operators $\smash{\hat A_i}$ which are differential
(i.e., local w.r.t.~$D_x$),
otherwise the above relations must be extended to a nonlocal setting.
The operators~$\smash{\hat A_i}$ are not always
compatible;\ a pair of Hamiltonian differential operators may generate
a nonlocal sequence~$\gA$. Further, we discuss the locality aspects
in more details.
\end{rem}

Consider the sequence $\gB\subset\sym\cEEL'$ of Noether symmetries~$\vph_i$
proliferated by the recursion
$R'=\square\circ A_1\circ\square^*\circ\ell_{\bgm}^{\bar u}$ such that the
integrals~$w$ define the Miura map $\gB\to\gA$ to a bi\/-\/Hamiltonian
sequence of Noether symmetries of the multi\/-\/component wave
equation~$\cE$.
The sequences $\gA$ and $\gB$ share the Hamiltonians, and
the second Hamiltonian structure $\smash{\hat A_2}$ for $\gA$ is
induced by the first Hamiltonian structure
$\smash{\hat B_1=(\ell_\bgm^{\bar u})^*}$ for~$\gB$ according to the
following diagram~(\cite{Magri}):
\begin{equation}\label{Diag}
\begin{diagram}
\cH_0 &&&& \cH_1 &&&& \cH_2 && {} \\
\dto_{\bE_\bu} & \rdto^{\bE_\gm} &&&
  \dto_{\bE_\bu} & \rdto^{\bE_\gm} &&&
  \dto_{\bE_\bu} & \rdto^{\bE_\gm} & {} \\
\Phi_0 & \pile{\lto^{\hat A_1} \\ \rto_{A_1}}
 & \phi_0 & \rto^{\hat A_2} &
 \Phi_1 & \pile{\lto^{\hat A_1} \\ \rto_{A_1}}
 & \phi_1 & \rto^{\hat A_2} &
 \Phi_2 & \pile{\lto^{\hat A_1} \\ \rto_{A_1}}  & \phi_2 \\
&& \dto^{\square} & & \uto_{\ell_\gm^\bgm} &&
 \dto^{\square} && \uto_{\ell_\gm^\bgm} &&
 \dto_{\square}  \\
&& \vph_1 & \pile{\rto^{\hat B_1} \\ \lto_{B_1}}
 & \psi_1 & \rto_{B_2}  &
 \vph_2 & \pile{\rto^{\hat B_1} \\ \lto_{B_1}}
 & \psi_2 & \rto_{B_2} & {\ldots}   \\
&&&\luto_{\bE_\bgm} & \uto_{\bE_{\bar u}} &&&
\luto_{\bE_\bgm} & \uto_{\bE_{\bar u}} \\
&&&& \cH_0 &&&&\cH_1 && {}
\end{diagram}
\end{equation}
The Magri schemes~(\cite{Magri}) for~$\gA$ and $\gB$ and
factorization~\eqref{eq4parts}, see~\cite{Miura},
of~$\smash{\hat A_2}$ are indeed standard.\ %
Now we see that the \emph{first} Hamiltonian structures~$\smash{\hat A_1}$
and $\hat B_1$ originate from the Lagrangians for the ambient
Euler\/-\/type equations and the
operator $\square$ 
is constructed
by using the integrals of the Liouvillean systems~$\cEEL'$.

If the Hamiltonian operators
$\smash{\hat A_1}=(\ell_\gm^u)^*$ and $\smash{\hat A_2}$,
see~\eqref{eq4parts}, are compatible, then the functionals $\cH_i[\gm]$
are in involution w.r.t.\ both structures,
the flows~$\phi_i$ commute (\cite{Magri}),
and~$\gA$ is infinite\/-\/Hamiltonian by Remark~\ref{RemInfinity}.
We also have
\begin{state}\label{WeaklyNonlocal}
Let a recursion~$R$ for an evolution equation
$\varepsilon=\{\bu_t=\phi[\bu]\}$ be constructed by using a single layer
of Abelian nonlocal variables
\textup{(}that is, expressions which are nonlocal w.r.t.\ the variables
in~$\varepsilon$ and which are produced by a set of
conservation laws~$\eta_\alpha$ for~$\varepsilon$,
see~\textup{\cite{ClassSymEng}}\textup{).}
Then~$R$ is \emph{weakly nonlocal}\textup{:}
$R=\text{differential part }+
   \sum_\alpha \vph_\alpha\,D_x^{-1}\circ\psi_\alpha$,
where $\vph_\alpha\in\sym\varepsilon$ and $\psi_\alpha$ is
the generating section of the conservation law~$\eta_\alpha$
for any~$\alpha$.
\end{state}

\noindent%
The proof of Proposition~\ref{WeaklyNonlocal} follows from the definitions;
it is generalized easily to the case of multiple layers of
the Abelian nonlocal variables defined by using (nonlocal) conservation laws.
Hence we conclude that the recursion
$R^*\colon\Phi_i\mapsto\Phi_{i+1}$ for the hierarchy~$\gA$
is weakly nonlocal owing to the factorization
$R^*=\smash{\hat A_2}\circ\smash{\hat A_1^{-1}}$ and Eq.~\eqref{eq4parts}.

\begin{state}
If   
the operators $\smash{\hat A_1}$ and $\smash{\hat A_2}$ are compatible,
then the flows $\phi_i\in\sym\cE$ are Noether.
\end{state}

\noindent%
Indeed, the homological formulation~(e.g.,~\cite{Lstar}) of the Magri
scheme and the triviality~(\cite{Getzler}) of the Poisson cohomologies
w.r.t.\ the structure~$\smash{\hat A_1}$ guarantee the existence
of the conservation laws~$\cH_i$ for~$\cE$;
now refer Lemma~\ref{MeshkovGenF} and the Noether
theorem~(\cite{Mystique, TodaLawsActa}).
We also conclude that~$\gA$ is local in~$w$.

Further, we demonstrate which of the above properties
can be transferred on~$\gB$.
Obviously, $\gA$ and $\gB$ are local simultaneously.
The restrictions of the mappings $B_2\colon\psi_i\mapsto\vph_{i+1}$ and
$\hat B_2\colon\vph_i\mapsto\psi_{i+1}$ onto the functionals
$H[w]\,\Id x$ that depend on $\bar u$ through the substitution
$w=w[\bgm]$ are Hamiltonian operators, and the functionals
$\cH_i[w[\bgm]]$ are in involution 
w.r.t.\ $\smash{\hat B_2}$ since it is so for~$\smash{\hat A_3}$;
from Remark~\ref{RemInfinity} we see why the Jacobi identity
holds for the operators~$B_2$ and $\smash{\hat B_2}$
if the Hamiltonians~$\cH$ depend on~$\bar u$ arbitrarily.

\begin{state}\label{InduceCommut}
Let the above notation and assumptions hold.
If the flows $\phi_i\in\gA$ commute and $\ker\smash{\hat A_2}=0$,
then the sequence~$\gB$ is also commutative.
\end{state}
\begin{proof}
By Proposition~\ref{NoetherForm},
every Noether symmetry of a Liouvillean Euler
hyperbolic equation $\cEEL'$ factors by the operator $\square$ (up to
the involution $x\leftrightarrow y$ if it is a symmetry of
$\cEEL'$).  The commutator of two Noether symmetries $\square(\phi')$,
$\square(\phi'')$ is a Noether symmetry again:
$\{\square(\phi')$, $\square(\phi'')\}=\square(\phi_{\{1,2\}})$,
although the structure of $\phi_{\{1,2\}}$ may contain addends
which are unusual w.r.t.\ the Jacobi bracket of $\phi'$ and $\phi''$,
see Eq.~\eqref{CommuteUnderSquare} below.
From Eq.~\eqref{eq4parts} it follows that
$\smash{\cEv_{\square(\phi_{\{1,2\}})}(w)=\smash{\hat A_2}(\phi_{\{1,2\}})}$.
Hence if the operator $\smash{\hat A_2}$ is injective,
the hierarchy $\gA$ is commutative, and the sequence $\gB$ is
correlated with $\gA$ by Eq.~\eqref{eq4parts} and~\eqref{Diag}, then
$\gB$ is 
commutative.
\end{proof}

From Proposition~\ref{InduceCommut} it follows that the recursion~$R'$
for~$\cEEL'$ is shared by all flows in~$\gB$.
By Proposition~\ref{WeaklyNonlocal}, the recursion
operator~$\smash{(R')^*}$ for the nonpotential flows in~$\gB$
is weakly nonlocal.

In Sec.~\ref{SecKdV} and \ref{SecBous}, we illustrate the scheme
above by using the multi\/-\/component modified
KdV and modified Boussinesq hierarchies: First, we
reconstruct the potential counterparts of $\gA$ treating the first
Hamiltonian structures $\smash{\Hat A_1}$ as the adjoint linearizations
of~\eqref{Constraint} and obtain the ambient hyperbolic Euler equations
by specifying their Lagrangians (that is, the matrices~$\bar\kappa$
since the Hamiltonians are trivial).
Then we find the Liouvillean Euler\/-\/type equations $\cEEL'$
such that the modified hierarchies $\gB$ are composed by
their symmetry flows and the Miura transformation
$\cEEL'\mapsto\cE$ that maps $\gB$ to $\gA$
is defined by the integrals of~$\cEEL'$.
Finally, we transfer the commutativity on
the local modified Noether flows
and factorize the weakly nonlocal recursions.

\section{2D Toda lattice and the KdV hierarchy}\label{SecKdV}
First, we recall some properties of the 2D Toda lattice (2DTL).
Let $r\geq1$ and suppose $K=\|k_{ij}\|$ is an arbitrary
real, constant, nondegenerate $(r\times r)$-matrix and
$\vec{a}={}^t(a_1$, $\dots$, $a_r)$ is a vector
such that $a_i\not=0$ for $1\leq i\leq r$
and the symmetry condition $\kappa_{ij}\equiv a_i k_{ij}=\kappa_{ji}$
holds for elements of the matrix $\kappa=\|\kappa_{ij}\|$. Then
the matrix $K$ is \emph{symmetrizable};
denote its inverse
by $K^{-1}=\|k^{ij}\|$. The $r$-com\-po\-nent two\/--\/dimensional Toda
lattice associated with the nondegenerate symmetrizable matrix $K$ is
$\cE_\Toda=\{u_{xy}=\exp(Ku)\}$.
The density $L_\Toda$ of its Lagrangian~$\cL_\Toda$ is
$L_\Toda=-\tfrac{1}{2}\langle u_y$, $\kappa u_x\rangle-\langle\vec{a}$,
$\exp(Ku)\rangle$; denote by $\vth=\kappa u_x$
the momenta obtained from~\eqref{Constraint}.
The component $w=\tfrac{1}{2}\langle
u_x$, $\kappa u_x\rangle-\langle\vec{a}$, $u_{xx}\rangle$ of the
energy\/--\/momentum tensor for $\cE_\Toda$ vanishes w.r.t.\ the
restriction $\bar D_y$ of the total derivative $D_y$ on $\cE_\Toda$:
$\bar D_y(w)=0$. The conformal weights $\smash{\vec\Delta}={}^t(\Delta^1$,
$\dots$, $\Delta^r)$ of the fields $\exp(u)$ are
$\Delta^i=\smash{\sum_{j=1}^r k^{ij}}$.
Consider the case when $w$ and its
differential consequences $\bw$ are unique solutions which depend on
the derivatives of $u$ to the 
equation $\bar D_y(\Omega)=0$; then we say that $K$ is \emph{generic}.

The Noether symmetries of $\cE_\Toda$ associated with a generic
symmetrizable $(r\times r)$-matrix $K$ are
$\vph=\square(\bE_w(H(x, \bw)\,\Id x))$
up to the transformation $x\leftrightarrow y$
(\cite{TodaLawsActa}, see also \cite{MeshkovTMPh}).
Here the
vector\/--\/valued, first\/--\/order differential operator $\square$ is
$\square=(\ell_w^\vth)^*=u_x+\smash{\vec\Delta}\,\bar D_x$
and $H$ is a smooth function.
The operator $R_\Toda=\square\circ\ell_s$, where $s_x=w$ and $s_{xy}=0$,
is a nonlocal recursion operator for the algebra $\sym\cE_\Toda$
(\cite{TodaLawsActa}).

Set $\phi_{-1}=1$ and generate three symmetry sequences for $i\ge0$:
$\vph_i=\square(\phi_{\scriptstyle i-1})\in\sym\cE_\Toda$,
$\phi_i=R_\pKdV(\phi_{i-1})\in\sym\cE_\pKdV$,
and $\Phi_i=\cEv_{\phi_i}(w)\in\sym\cE_\KdV$,
where the KdV equations are~\eqref{BothKdV} and $\beta\equiv\langle
\vec{a},\vec{\Delta}\rangle$.
The evolutions $\Phi_i$ and $\phi_i$ are elements of the local
commutative bi\/--\/Hamiltonian hierarchies for the (p)KdV equations
$w_{t_1}=\Phi_1$ and $s_{t_1}=\phi_1$, respectively.
The equation $u_{t_1}=\vph_1=\square(\phi_0)$
is an $r$-component analogue of the potential modified KdV equation,
being a Noether symmetry of the Toda lattice~$\cE_\Toda$ as well.
Equation $u_{t_1}=\vph_1$ is Hamiltonian
w.r.t.\ the operator $B_1=D_x^{-1}\,\kappa^{-1}$.
The $r$-component analogue $\cE_\mKdV=\{\vth_{t_1}=\kappa\,D_x(\vph_1)\}$
of the modified KdV equation
is Hamiltonian w.r.t.\ $\Hat B_1=\kappa\,D_x$.
The integral~$w[\vth]$ defines a Miura transformation between the
higher modified KdV
equations and the hierarchy~$\gA$ of Eq.~\eqref{BothKdV}.
Indeed, relation~\eqref{eq4parts} holds for the transformation
$w=w[\vth]$ and the Hamiltonian operator
$\smash{\hat A_2}=-\beta D_x^3+D_x\circ w+w\cdot D_x$
for the KdV equation~$\cE_\KdV$. Hence,
the times $t_i$ of the evolutions $u_{t_i}=\vph_i$ are correlated with
the times in the hierarchy $\gA$ for $\cE_\pKdV$ by this map,
and the four equations $\cE_{\text{\textup{(p)(m)KdV}}}$
and their higher analogues as well share the same set of the
Hamiltonians~$\cH_i=[H_i\,\Id x]$.


\begin{state}[\textup{\cite{DynamSys}}]
The factorizations $R_\Toda=\square\circ\ell_s$ and
$R_\pKdV=\ell_s\circ\square$ hold for the recursions
$R_\Toda\colon\vph_i\mapsto\vph_{i+1}$ and
$R_\pKdV\colon\phi_i\mapsto\phi_{i+1}$.
Every flow $\vph_k=\square\,\circ\,\bE_w(\cH_{k-1})\in\sym\cL_\Toda$
is a Noether symmetry of the Toda equation, and $\vph_k$ is
Hamiltonian w.r.t.\ the Hamiltonian structure
$B_1=\kappa^{-1}\cdot D_x^{-1}$ and the Hamiltonian
$\cH_{k-1}=[H_{k-1}\,\Id x]$ for the $(k-1)$\textup{th}
higher KdV~
equation.
\end{state}

\begin{state}\label{mKdVCommute}
The symmetries $\vph_k$ commute.
\end{state}

\begin{proof}
The KdV hierarchy $\gA$ is commutative
and local in $s_x=w$.
Let $\vph'=\square\,(\phi'(x$, $\bw))$ and
$\vph''=\square\,(\phi''(x$, $\bw))$,
then the Jacobi bracket of $\vph'$ and $\vph''$ is
$\{\vph'$, $\vph''\}=\square\,(\phi_{\{1,2\}}(x$, $\bw))$, where
(see~\cite{SokolovUMN} for the scalar case)
\begin{equation}\label{CommuteUnderSquare}
\phi_{\{1,2\}}=\cEv_{\vph'}(\phi'')-\cEv_{\vph''}(\phi')+\bar
D_x(\phi')\,\phi''-\phi'\,\bar D_x(\phi'').
\end{equation}
We have $[\cEv_{\vph_{k_1}}$, $\cEv_{\vph_{k_2}}](w)=\{\Phi_{k_1}$,
$\Phi_{k_2}\}=0$. We also have
$\vph_{k_i}=\square(\phi_{k_{\scriptstyle i-1}})$ and
$\cEv_{\{\vph_{k_1},\vph_{k_2}\}}(w)=\Hat A_2(\phi_{\{k_1,k_2\}})$.
%
Obviously, the Hamiltonian operator~$\Hat A_2$ for KdV is injective.
Therefore, $\phi_{\{k_1,k_2\}}=0$ and $\{\vph_{k_1}$,
$\vph_{k_2}\}=\square(0)=0$.
We also conclude that
the operator $R_\Toda$ is a recursion for the $i$th equation
$u_{t_i}=\vph_i$ within the commutative symmetry subalgebra $\gB$,
here $i\geq0$ is arbitrary.
\end{proof}

\section{The Boussinesq hierarchy}\label{SecBous}
The first Hamiltonian structure
for the Boussinesq equation~\eqref{BothBous} is
$\smash{\hat A_1}=\left(
\begin{smallmatrix} 0 & D_x \\ D_x & 0 \end{smallmatrix}\right)$.
Hence $u$, $v$ such that $U=v_x$, $V=u_x$ are the 
potentials
satisfying the potential Boussinesq equation, see Eq.~\eqref{BothBous}.
Further on, we use the notation $\bu={}^t(u,v)$ and
$\bU={}^t(U,V)$.
The Lagrangian functional with the density
$L_\Bous=-\tfrac{1}{2}u_xv_y-\tfrac{1}{2}v_xu_y$ for the ambient
two\/-\/component wave equation $\cE=\{v_{xy}=0$, $u_{xy}=0\}$ is
obtained straighforwardly. A 
Miura transformation for Eq.~\eqref{BothBous}
is given by the formulas~(\cite{PavlovFPM})
\begin{equation}\label{IntegralsmBous}
U=a^2+ab+b^2+2a_x+b_x,\qquad
V=-2a(b(a+b)+b_x)-D_x(b_x+ab+b^2)
\end{equation}
such that the nonpotential modified Boussinesq equation is
\begin{equation}\label{mBous}
a_t=\tfrac{1}{3}D_x(a^2-2ab-2b^2-a_x-2b_x),\quad
b_t=\tfrac{1}{3}D_x(-2a^2-2ab+b^2+2a_x+b_x).
\end{equation}
Equation~\eqref{mBous} is Hamiltonian w.r.t.\ the Hamiltonian
operator $\hat B_1=-\smash{\tfrac{1}{3}\left(\begin{smallmatrix}
\pp2 & -1 \\ -1 & \pp2\end{smallmatrix}\right)}\cdot D_x$ and the
Casimir $\cH_0=[V\,\Id x]$. (We note that $\hat B_1$
contains the Cartan matrix~$K_{\mathfrak{sl}_3}$
of the algebra~$\mathfrak{sl}_3(\BBC)$,
see~\cite{FordyGibbons}. Also, recall that the second sequence of flows
for Eq.~\eqref{BothBous} starts with the Casimir $\bar\cH_0=[U\,\Id
x]$. Now we see that the integral constraints
$[U\,\Id x]=\int_{-\infty}^{+\infty} v_x\,\Id x=\text{const}$,
$[V\,\Id x]=\int_{-\infty}^{+\infty} u_x\,\Id x=\text{const}$
used in the inverse scattering problem method are natural.
Moreover, note that the Hamiltonians $\cH_0$ and $\bar\cH_0$
define \emph{nonzero} symmetry flows on Eq.~\eqref{BothBous},
which are the shifts of the potential variables $u$ and $v$, respectively.)
Therefore, we introduce the 
potentials $\alpha$, $\beta$ such
that $a=\tfrac{1}{3}(2\alpha_x-\beta_x)$,
$b=\tfrac{1}{3}(-\alpha_x+2\beta_x)$
and denote $\balpha={}^t(\alpha,\beta)$ and $\ba={}^t(a,b)$.
The Hamiltonian flows ${\bu}_{t_{i-1}}=\phi_{i-1}$ are mapped
to the flows ${\balpha}_{t_i}=\vph_i=\square(\phi_{i-1})$
by the operator
\[
\square=(\ell_{\bU}^{\ba})^*=\begin{pmatrix}
2a+b-2D_x & -4ab-2b^2-2b_x+b\,D_x \\
a+2b-D_x & -2a^2-4ab+2D_x\circ a-D_x^2+(a+2b)D_x)
\end{pmatrix}.
\]
The ambient Euler equation $\cEEL'$ is described by solving the
equations $\{\vph_i\in\sym\cEEL'\}$, $i\geq0$, w.r.t.\ the Hamiltonian
$H'_{\text{EL}}$; solving the first two, we obtain
$H_{\text{EL}}^\prime=
c_1\exp(\beta-\alpha)+c_2\exp(\alpha)+c_3\exp(-\beta)$,
where $c_1$, $c_2$, $c_3\in\BBR$, and hence we get
the 2DTL. The functionals $\bU[\ba]$ are
integrals for $\cEEL'$ (hence it is Liouvillean) if $c_2=0$.
Thus we obtain the system
\begin{equation}\label{TodaFixedPoint}
\alpha_{xy}=-c_1\exp(\beta-\alpha)   
            -c_3\exp(-\beta),\quad
\beta_{xy}=c_1\exp(\beta-\alpha)     
            -2c_3\exp(-\beta),
\end{equation}
which is transformed to the 2DTL~(\cite{FordyGibbons})
$\balpha_{xy}=\exp(K_{\mathfrak{sl}_3}\balpha)$
by a change of variables.
Recall that $\phi_i\in\mathrm{im}\,\bE_{\bU}$ and
$\square=(\ell_{\bU}^{\ba})^*$. Therefore the flows
$\vph_i=\square(\phi_{i-1})$ are Noether symmetries of~$\cEEL'$.
Then the Miura transformation $\bU=\bU[\ba]$ and Eq.~\eqref{eq4parts}
provide the natural factorization of the
second Hamiltonian structure~(\cite{Olver})
\begin{equation}\label{BousSecondHam}
\hat A_2=\begin{pmatrix}
-2D_x^3+2U\,D_x+U_x & {3}V\,D_x+2V_x \\
{3}V\,D_x+V_x &
\begin{gathered}
{} \\
\tfrac{2}{3}D_x^5-\tfrac{5}{3}(U\,D_x^3+D_x^3\circ U)+{}\\
{}+U_{xx}\,D_x+D_x\circ U_{xx}+\tfrac{8}{3}U\,D_x\circ U
\end{gathered}
\end{pmatrix}
\end{equation}
for Boussinesq equation~\eqref{BothBous}.
The initial terms of the Boussinesq hierarchy and the
correlated sequence of modified Boussinesq flows that share the
Hamiltonians $\cH_i$ are shown in diagram~\eqref{Diag}.
Here the Hamiltonian densities are $H_0=V$,
$H_1=\tfrac{1}{2}V^2+\tfrac{1}{6}U_x^2+\tfrac{2}{9}U^3$, and
$H_2=\tfrac{1}{12}U_{xx}V_{xx}-\tfrac{5}{12}UU_{xx}V-\tfrac{5}{16}U_x^2V+
\tfrac{5}{36}U^3V+\tfrac{5}{48}V^3$, etc.
From Proposition~\ref{InduceCommut} we
conclude that the elements $\vph_i$ of the sequence~$\gB$ of
modified Boussinesq Noether flows on 2DTL~\eqref{TodaFixedPoint} are
local in~$\ba$ and pairwise commute.
The proof is analogous to Proposition~\ref{mKdVCommute};
we note that the explicit formula, Eq.~\eqref{CommuteUnderSquare},
for the bracket~$\phi_{\{k_i,k_j\}}$ is not used and hence
may be not known. Finally, we obtain the factorizations
$R_{\mathstrut\text{pmBous}}=
\square\circ\smash{\hat A_1^{-1}}\circ\square^*\circ\ell_{\ba}^{\balpha}$ and
$R_{\mathstrut\text{mBous}}=
\ell_{\ba}^{\balpha}\circ\square\circ\smash{\hat A_1^{-1}}
\circ\square^*$ for the recursion operators
$\smash{R_{\mathstrut\text{pmBous}}^{\mathstrut}=
R_{\mathstrut\text{mBous}}^*}$
that proliferate higher modified
Boussinesq equations~\eqref{mBous}. By Proposition~\ref{WeaklyNonlocal},
the recursions for~$\gA$ and~$\gB$ are weakly nonlocal.

\subsection*{Acknowledgements.}
The author thanks B.~A.~Dubrovin, I.~S.~Kra\-sil'\-sh\-chik,
M.~V.~Pavlov, and
A.~M.~Verbovetsky for useful discussions and is profoundly grateful
to V.~V.~Sokolov and the anonymous referee
for remarks and constructive criticism.
The author thanks the organizing
committee of the International Workshop ``Nonlinear Physics: Theory and
Experiment~III'' for support. The work is partially supported by the
INTAS\- grant YS\,2001/2-33 and the University of Lecce grant
No.\,650\,CP/D. The author is grateful to the University of Lecce
and Brock University for warm hospitality.


\begin{thebibliography}{77}\normalsize

\bibitem{ClassSymEng}
\by{Bocharov A.\,V., Chetverikov V.\,N., Duzhin S.\,V., et al.}
\book{Symmetries and conservation laws for differential equations of
mathematical physics}.\ AMS, Providence, RI, 1999.\ %
I.~Krasil'shchik and A.~Vinogradov, eds.

\bibitem{Demskoi}
\by{Demskoi D. K., Startsev S. Ya.}
// \jour{Fundam.\ Appl.\ Math.} \vol{10} (2004), n.1, 29--37.


\bibitem{SokolovUMN}
\by{Zhiber A. V., Sokolov V. V.}
// \jour{Russ.\ Math.\ Surveys} \vol{56} (2001), n.1, 61--101.

\bibitem{LeznovSmirnovShabat}
\by{Leznov A. N., Smirnov V. G., Shabat A. B.}
// \jour{Teor.\ matem.\ fizika} \vol{51} (1982), n.1, 10--21.

\bibitem{MeshkovTMPh}
\by{Meshkov A. G.}
// \jour{Teor.\ matem.\ fizika} \vol{63} (1985), n.3, 323--332.

\bibitem{PavlovFPM}
\by{Pavlov M. V.}
// \jour{Fundam.\ Appl.\ Math.} \vol{10} (2004), n.1, 175--182.

\bibitem{ShabatYamilov}
\by{Shabat A. B., Yamilov R. I.} Exponential systems of type~I
and the Cartan matrices. Preprint. Ufa, 
1981. 

\bibitem{Mystique}
\by{Barnich G., Brandt F., Henneaux M.}
// \jour{Commun.\ Math.\ Phys.} \vol{174} (1995), 57--92.

\bibitem{FordyGibbons}
\by{Fordy A. P, Gibbons J.}
// \jour{J.~Math.\ Phys.} \vol{21} (1980), n.10,
2508--2510;
\jour{J.~Math.\ Phys.} \vol{22} (1981), n.6, 1170--1175.

\bibitem{Gardner}
\by{Gardner C. S.}
// \jour{J.~Math.\ Phys.} \vol{12} (1971), n.8, 1548--1551.

\bibitem{Getzler}
\by{Getzler E.}
// \jour{Duke Math.~J.} \vol{111} (2002), 535--560.


\bibitem{Lstar}
\by{Kersten P., Krasil'shchik I., Verbovetsky A.}
// \jour{J.~Geom.\ Phys}. \vol{50} (2004), n.1-4, 273--302.

\bibitem{TodaLawsActa}
\by{Kiselev A. V.} 
// \jour{Acta Appl.\ Math.} \vol{83} (2004), n.1-2,
175--182.

\bibitem{DynamSys}
\by{Kiselev A. V., Ovchinnikov A. V.}
// \jour{J.~Dynamical and Control Systems} \vol{10} (2004), n.3, 431--451.

\bibitem{Magri}
\by{Magri F.}
// \jour{J.~Math.\ Phys.} \vol{19} (1978), n.5, 1156--1162.

\bibitem{Miura}
\by{Miura R. M.}
// \jour{J.~Math.\ Phys.} \vol{9} (1968), 1202--1204.

\bibitem{NutkuPavlov}
\by{Nutku Y., Pavlov M. V.}
// \jour{J.~Math.\ Phys.} \vol{43} (2002), n.3, 1441--1459.

\bibitem{Olver}
\by{Olver P. J.} \book{Applications of Lie groups to differential
equations}, $2$nd ed.,  
Springer\/--\/Verlag, NY, 1993.



\end{thebibliography}
\end{document}